\input{aipcheck}

\documentclass[
    ,final            
  ]
  {aipproc}

\layoutstyle{6x9}

\def\simless{\mathbin{\lower 3pt\hbox
   {$\rlap{\raise 5pt\hbox{$\char'074$}}\mathchar"7218$}}} 
\def\simgreat{\mathbin{\lower 3pt\hbox
   {$\rlap{\raise 5pt\hbox{$\char'076$}}\mathchar"7218$}}} 

\def\kms{{\rm\,km\,s^{-1}}}

\def\gcm3{{\rm g\,\, cm^{-3}}}
\def\msun{M_\odot}
\def\pc3{{\rm pc}^{-3}}
\def\ergs{\rm {erg\,\,s^{-1}}}

\begin{document}

\title{THE DYNAMICAL FINGERPRINT \break OF INTERMEDIATE MASS \break BLACK HOLES
\break IN GLOBULAR CLUSTERS}

\classification{97.60.Ls;  97.60.Jd;  97.80.Jp}
\keywords      {Black holes; pulsars; binaries; globular clusters}

\author{M. Colpi}{
  address={Department of Physics G. Occhialini, 
University of Milano Bicocca, Piazza
della Scienza 3, Milano, I 20126 Italy}
}
\author{B. Devecchi}{
  address={Department of Physics G. Occhialini, 
University of Milano Bicocca, Piazza
della Scienza 3, Milano, I 20126 Italy}
}
\author{M.  Mapelli}{
  address={SISSA/ISAS, 
Via Beirut 4, Trieste, I 34014 Italy}
}
\author{A. Patruno}{
  address={Department of Physics G. Occhialini, 
University of Milano Bicocca, Piazza
della Scienza 3, Milano, I 20126 Italy}
}
\author{A. Possenti}{
  address={INAF, Cagliari Observatory, 
Poggio dei Pini, Strada 54, Cagliari, I 09012 Italy}
}

\begin{abstract}
A number of observations hints for the presence of an intermediate
mass black hole (IMBH) in the core of three globular clusters: M15 and NGC
6752 in the Milky Way, and G1, in M31.  However the existence of 
these IMBHs 
is far form being conclusive.  In this paper, we review their main formation
channels and explore possible observational signs that a single or
binary IMBH can imprint on cluster stars. In particular we explore the role played by a binary IMBH
in transferring angular momentum and energy to stars flying by.

\end{abstract}

\maketitle

\section{Introduction}
A number of different observations suggest that large black holes
(BHs) may exist in nature, with masses between $20\msun - 10^4
M_{\odot}.$ Heavier than the stellar-mass BHs born in core-collapse
supernovae ($3\msun-20\msun$; \cite {Oroz}), these intermediate mass
black holes (IMBHs) are expected to form in dense stellar systems
through complex dynamical processes. 
Globular clusters thus become
prime sites for their search. Recently, Gebhardt, Rich, \& Ho
\cite{GRH02} suggested the presence of an IMBH of
$2^{+1.4}_{-0.8}\times 10^4\msun$ in the globular cluster G1, in M31,
to explain its kinematics and surface brightness profile.  Gressen et
al. \cite{GRE03} indicate the presence of
an IMBH of $1.7^{+2.7}_{-1.7}\times 10^3\msun$ in the galactic globular
cluster M15, based on HST kinematical data.  An additional puzzling
observation comes from the exploration of the globular cluster NGC
6752 with the discovery of 5 millisecond pulsars (MSPs) showing
unusual accelerations or locations
\cite{DAPO02}.

NGC 6752 hosts in its core two MSPs (PSR-B and PSR-E) with very high
negative spin derivatives that, once ascribed to the overall effect of
the cluster potential well, indicate the presence of $\sim 1000\msun$
of under-luminous matter enclosed within the central 0.08pc
\cite{FER03a}.  NGC 6752 in addition hosts two MSPs with
unusual locations: PSR-A, a canonical binary pulsar with a white dwarf
companion \cite{DAPO02,FER03b}, holds the record of being the farthest MSP ever observed in
a globular cluster, at a distance of $\approx 3.3$ half mass radii,
and PSR-C, an isolated MSP that ranks second in the list of the most
offset pulsars known, being at a distance of 1.4 half mass radii from
the gravitational center of the cluster \cite{DAPO02}.  Colpi, Possenti
\& Gualandris \cite{CPG02} first conjectured that PSR-A was propelled
into the halo in a flyby between the binary MPS and a
binary stellar-mass BH or a binary IMBH present in the core of NGC
6752.  Colpi, Mapelli \& Possenti (CMP03 hereon; \cite{CMP03})
proposed later that the position of PSR-A could
also be explained as an ejection following a dynamical encounter of a
non-recycled neutron star in a binary, by a single 
IMBH, prompted by the evidence of under-luminous matter in
the core of NGC 6752 \cite{FER03a}.
The interaction considered was a flyby between the binary pulsar PSR-A
and the IMBH having within its sphere of influence a cusp star bound
on a Keplerian orbit.  CMP03 carried on an extensive analysis of
binary-binary encounters with IMBHs, single or in binaries,  
to asses the viability of their
scenario, indicating that IMBHs are best targets for imprinting the
necessary thrust to PSR-A at a rate compatible with its persistence in
the halo against dynamical friction.  Ejection of PSR-A from the core
to the halo following exchange interactions off binary stars can not be
excluded, but as pointed out by Colpi et al. \cite{CPG02}, the binary
parameters of PSR-A and its evolution make this possibility remote,
and call for fine tuning conditions on binary evolution.

All three these observations, hinting for an IMBH interpretation of the
data, are far from being conclusive as regard to the nature of their
dark component.  Numerical studies by Baumgardt et al. \cite{BAU03}
have in fact shown that
kinematical features observed in G1 and M15 can be explained if dark
low-mass remnants reside in their cores, without need of an exotic
IMBH.  Also for NGC 6752, the underluminous matter found can be
associated to a cluster of compact stars \cite{FER03a}.  
Thus, other signs of an IMBH should be explored in
order to asses the reliability of such interpretations.

On theoretical ground the existence of IMBHs in globular clusters,
single or in binaries, has been advanced by several authors (see van
der Marel for a review \cite{vMar04}), but the difficulty in finding a
clear formation pathway remains.  Recently, Portegies Zwart et
al. \cite{PZMc02,PZCO04} suggested that IMBHs find their formation
channel in young star clusters sufficiently dense to become vulnerable
to unstable mass segregation \cite{PZBHM04,Frai05}.  Through the
runaway collision of a single heavy star off other stars, a giant
stellar object is expected to form that collapses into an IMBH. If
this holds true in globular clusters at the time of their formation,
there is freedom to believe that gas-dynamical processes, as such
suspected to occur in young metal rich star clusters, were at work
early in the cluster lifetime.

It has also been speculated that 
IMBHs in globular clusters may form, 
alternatively, through binary-single or binary-binary 
gravitational encounters and 
mergers among light BHs during a far more advanced stage of cluster evolution
\cite{MH02}.
In clusters a few billions years old, the heaviest stars are
stellar compact remnants, i.e., neutron stars and black holes. 
Despite neutron stars form in larger
numbers (for any reasonable initial mass function), 
BHs likely outnumber neutron stars, since they experience
weaker natal kicks (due to their larger inertia) and are thus easily
retained inside the cluster.  The
end-result of stellar and dynamical
evolution is a dense core of stellar-mass
BHs, some bound to stars in binaries.  Sinking further by unstable mass
segregation these BHs decouple dynamically
from the system and get caught in binaries through exchange
interactions among BHs \cite{KUMc93,SH93}.  These binaries, in the high density
environment of a mass-segregated core, experience frequent
interactions, initiating a process of hardening that may
proceed until gravitational wave emission drives the evolution 
of the binaries toward coalescence
into a single more
massive BH. The process may repeat leading to a larger IMBH \cite{MH02}. 

The  hardening
of binaries via gravitational encounters can however find sudden
halt if the BH binary is light enough to experience ejection:
since the interactions that produce hardening produce recoil,
a sizable fraction of BHs can be ejected
so that the core of BH remnants evaporates and dissolves \cite{KUMc93,SH93,PZMc00}. 
Single as well as binary BHs leave the cluster, 
emptying the core of all its BHs but a small number.
Thus, it is from a delicate balance between hardening and recoil that
sequences of encounters among 
BHs can drive the core into a state with no BHs or with a few, 
bound preferentially in binaries. 
Miller \& Hamilton \cite{MH02} and CMP03 showed,
from simple considerations,  that a minimum 
initial seed mass is required (around 50$\msun$) for the BH to remain
in the cluster and grow up to several hundred solar masses, through
hardening and coalescence.
But a closer and more detailed inspection of sequences of
binary-single scattering events by Gultekin, Miller \& 
Hamilton \cite{Gult04} have revealed 
that in order to avoid ejection, a larger seed BH
should exists in situ of $\sim 300\msun$, 
at the onset of dynamical evolution to remain safely 
inside the cluster and grow further in mass, avoiding ejection.
There might be also the possibility that BHs propelled away from the
core remain bound to the cluster living in the halo. These BHs may return
back by dynamical friction after almost all the central BHs have
been expelled by recoil.
Since there is no unique outcome for the fate of BHs in globular
clusters, various scenarios remain open  for investigation.
A globular cluster may host:

\begin{itemize}
\item
a {\it single} IMBH, 
with mass $> 300\msun$ formed
in situ following a runaway merger among heavy stars, at the time
of cluster formation. This IMBH may subsequently
grow up to $10^{3}-10^{4}\msun$ or more, by dynamical
process,  on the core relaxation time. 
This IMBH may capture  stars via relaxation processes or
gravitational interactions and be surrounded  by a small cusp of stars.

\item 
a {\it binary} IMBH of mass $50\msun-300\msun$  with a
BH companion of similar mass or lighter then  $ 10\msun.$   This binary may
form dynamically via exchange interactions and mergers 
among BHs.    
The large cross section that such a binary has implies  
that it is relatively shortlived since close encounters with 
stars can cause its hardening up to coalescence.

\item a {\it stellar-mass binary} BH can be present composed of
two BHs, relic of the most massive stars, perhaps ejected in the
halo, that returned back to the core by dynamical friction after BH core 
evaporation.

\end{itemize}

\section {Single intermediate mass black holes}

If a single IMBH of mass $\simgreat 10^3\msun$ 
is present in a globular cluster, it can influence
stars passing by in various ways. 
A case of interest occurs when the IMBH is not strictly single, but 
is surrounded by a swarm of stars, i.e., a small cusp.
This cusp is likely to be unstable  to 
ejections,  captures and relaxation processes. It is however  not  
implausible that at least a tightly bound star is present. 
Given the high fraction of binary
stars in the core of globular clusters that form dynamically, the
interaction of these binaries with the IMBH may bring 
to three potential observational signatures:

\noindent
(i) {\it Flybys} involving the IMBH
and its cusp star. As discussed in CMP03, a flyby may
explain the ejection from the core into the cluster halo 
of binary stars, such as PSR-A. 

\noindent
(ii) {\it Ionization of the binaries}, 
either interacting with the IMBH and its cusp
star (CMP03) or experiencing the tidal field of the IMBH itself
\cite{pfahl04}.  As discussed recently by Pfahl \cite{pfahl04}, ionization 
by the tidal field can bring one
binary component into a bound-disruption orbit, releasing
the other on an hyperbolic orbit. The IMBH in this case may
reveal its presence flaring in X-ray when swallowing the tidal 
debris of the disrupted star every $\sim$1-10 Myr, 
depending on the details of the capture process.

\noindent
(iii) {\it Accretion.} Binary-single or binary-binary encounters with the
IMBH can deliver a star on a close orbit around the IMBH. After
circularization, a phase of mass transfer can initiate, similarly to
what happen in X-ray binaries hosting a stellar-mass BH.

\subsection {Accretion onto an IMBH}

We have explored accretion onto an IMBH considering a low mass donor
star, evolving along the main sequence and red giant branch.  The
donor star is modeled using an updated version of the evolutionary
code of Eggleton \cite{Eg71} and of Webbink, Rappaport \& Savonije \cite{Web83}. 
\begin{figure}
 \includegraphics[height=.3\textheight]{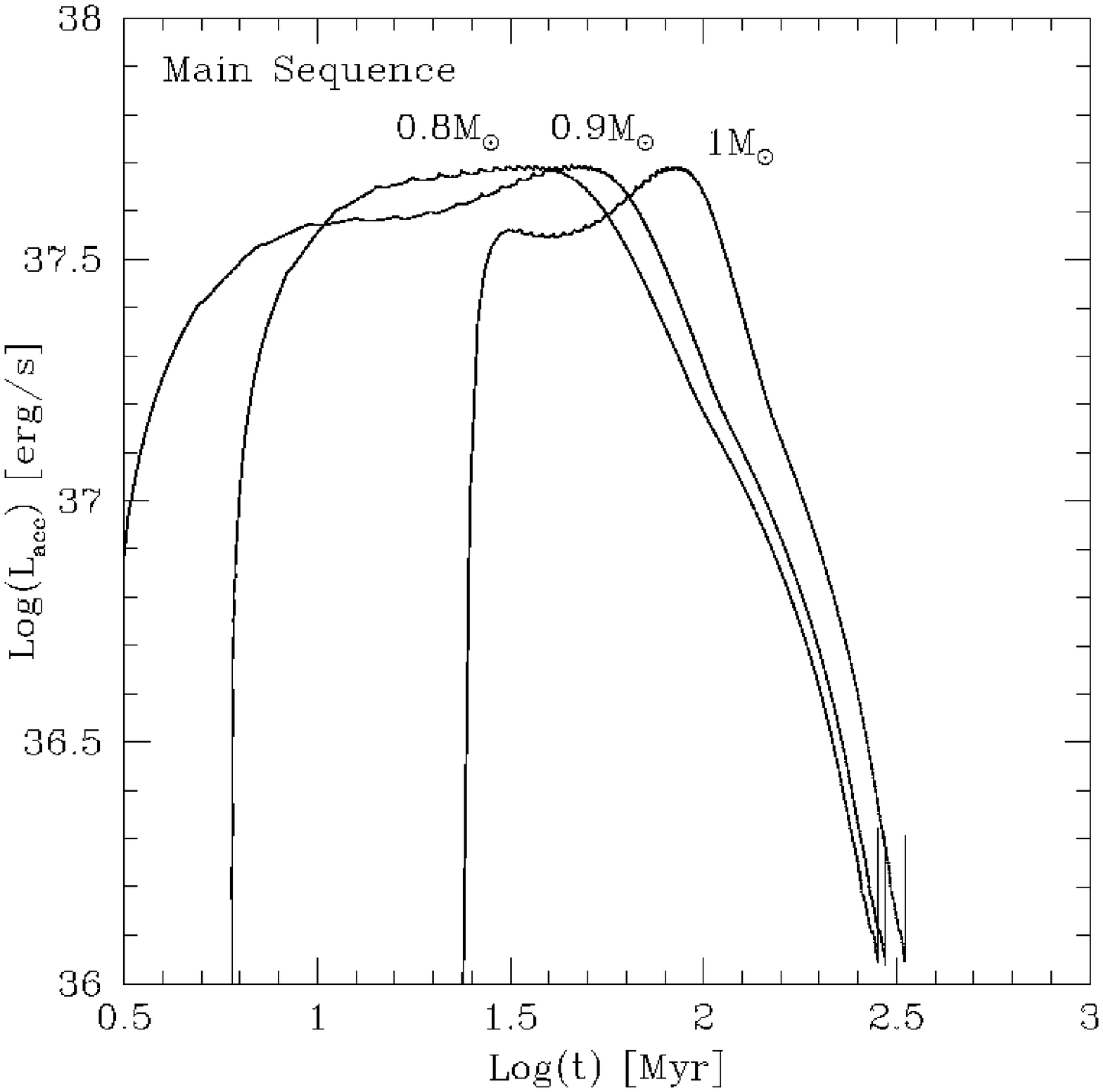}
  \includegraphics[height=.3\textheight]{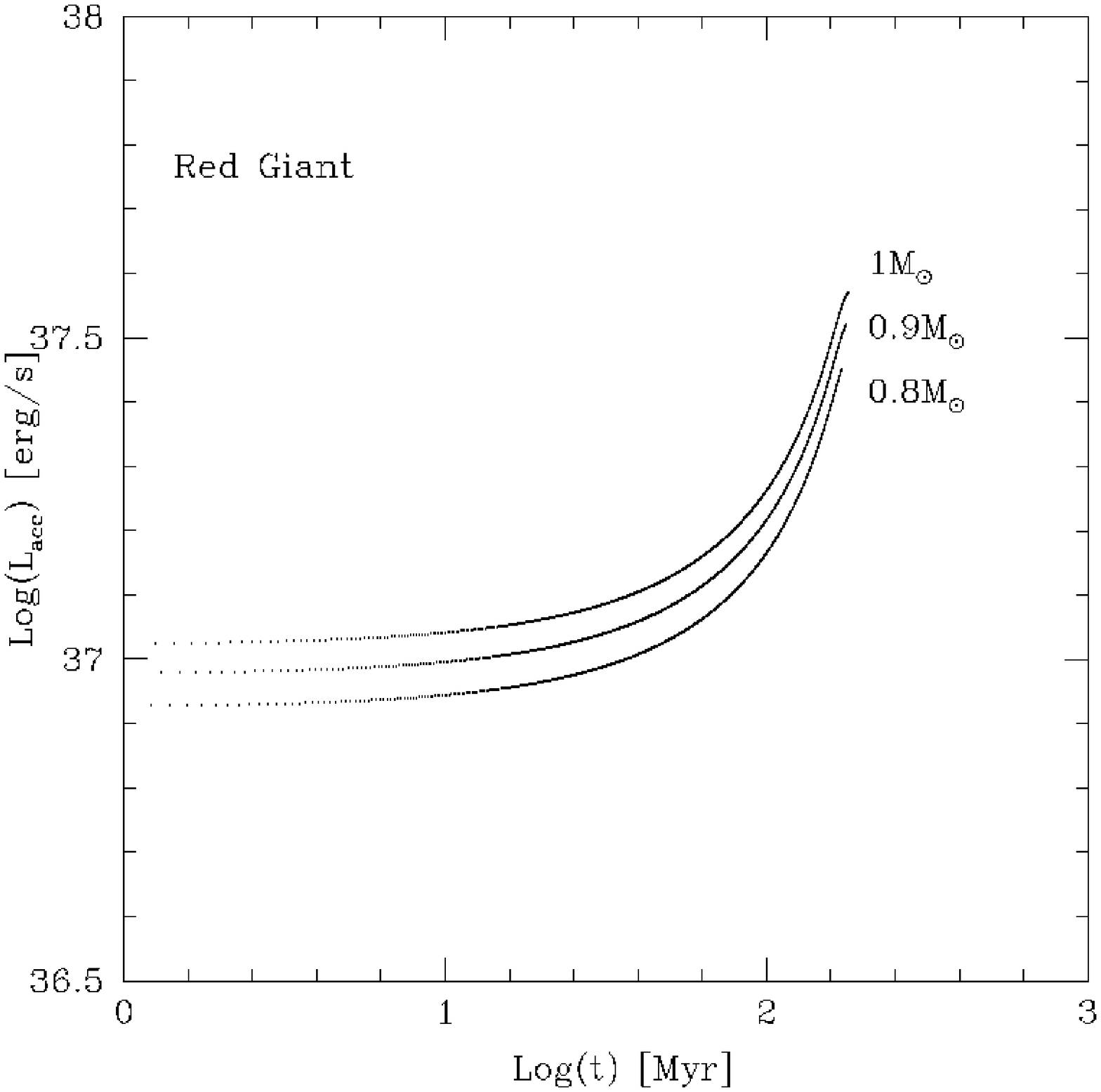}
  \caption{Luminosity versus time from accretion onto an 
IMBH of 1000$\msun.$ The donor is a light star of 0.8,0.9 and 1$\msun$, respectively. 
Left (right) panel
refers to accretion when the donor is on the main sequence (red giant phase).}
\end{figure}
Figure 1 shows the run of the luminosity versus
time: mass transfer via Roche lobe 
overflow leads, in both cases, to luminosities  $\simgreat 
10^{37}\ergs$.   These correspond to mean accretion rates 
that are low enough to fulfill the condition for variable mass transfer in 
a thermal-viscous unstable thin disk 
\cite{Dubus99}. Thus, we find that IMBHs accreting from low mass donors 
should undergo limit-cycle behavior and appear as transient X-ray sources.
The signs that would distinguish an IMBH from a stellar-mass BH
in a low-mass X-ray binary 
should thus be searched in the spectral and timing properties: 
a softer 
black body component and longer timescale variabilities  
may be the distinguishing features \cite{MFM04}.  
We can not exclude a priori that bright X-ray sources   
seen in ellipticals, having globular clusters as
optical counterparts, host accreting IMBHs \cite{FabWh04,Mu04}. 

\section {Binary intermediate mass  black holes}

A binary IMBH can imprint large recoil
velocities to stars flying by. When not too hard and massive, 
the binary IMBH can also  
transfer angular momentum to the stars, perturbing them 
away from dynamical equilibrium. Whether this effect influences only few stars
or a sizable number is 
still unexplored and under our current study (Mapelli et al. 2005 in 
preparation).
We here report preliminary results 
obtained running single-binary encounters
between a binary IMBH and cluster stars.

\subsection{Supra-thermal stars and angular momentum alignment?}

Our aim is to address the following questions: (i) 
Are stars heated to supra-thermal energies,
i.e. to energies in excess of their dispersion values 
without escaping from the globular clusters?
(ii) Is there direct transfer
of angular momentum from the binary IMBH to the stars?
(iii) Do we observe alignment of the stellar orbit 
in the direction of the angular momentum of the binary?
\begin{figure}
  \includegraphics[height=.35\textheight]{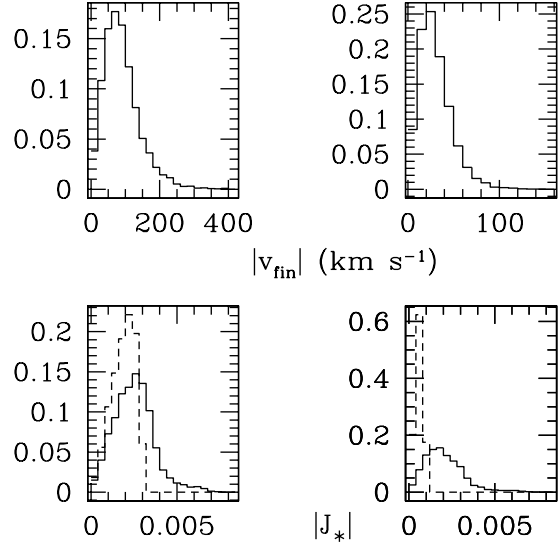}
  \caption{Post-encounter velocity and angular distribution (in modulus)
of cluster stars scattering off 
a binary IMBH of mass $[100\msun,50\msun],$ separation of 10 AU (left panels) and 100 AU
(right panel), 
and eccentricity $e=0.7$. }
\end{figure}

Figure 2 shows the post-encounter velocity and angular momentum
distributions (in modulus) of stars that impinge on a binary IMBH of mass
$[100\msun,50\msun]$, and orbital separation $a$ of 10,100 AU, respectively. 
The hard (softer) binary, with $a=10$ AU (100 AU), tends
to produce stars with velocities above (below
or closer to) the escape speed ($\simeq 40\kms$).
Respectively, 15\% and 80\% of the stars are scatterd to 
supra-thermal energies.
Angular momentum is exchanged when the binary IMBH 
has  a higher orbital angular momentum relative to
that of the incoming stars.
The right panel of Figure 3 shows the most favorable case of angular
momentum alignment (for the binary IMBH with $a=100$ AU): 
we find that  alignment involves
a sizable fraction of stars ($\sim70\%$), 
so introducing an anisotropy in their 
equilibrium energy and angular momentum distributions. 
Supra-thermal stars are also those absorbing the largest angular momentum.
\begin{figure}
  \includegraphics[height=.3\textheight]{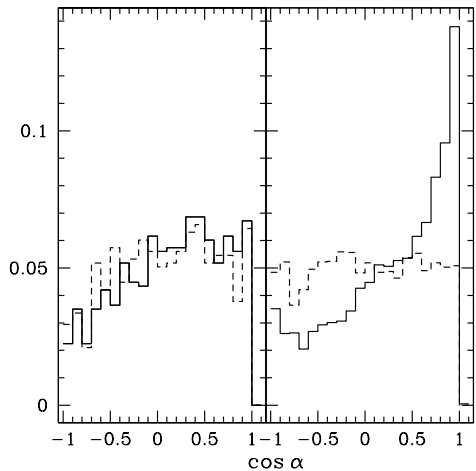}
  \caption{The distribution of the angle $ \alpha$
between the binary IMBH angular 
momentum and the orbital angular momentum of the incoming star. 
The binary IMBH is as in Figure 2. 
Solid (dashed) line indicates the post-encounter (pre-encounter) 
distributions.
Note that there is an overabundance of 
corotating stars when the binary IMBH has $a=100$ AU (right panel). }
\end{figure}

A binary IMBH tightens rapidly via binary-single flybys and the bulk
of the bound supra-thermal stars with excess angular momentum are
produced over a time comparable to the IMBH hardening life, typically
of $\sim 10^7-10^8$ yrs for a cluster such as NGC 6752.  Propelled
into the halo, these stars mainly return to equilibrium within a few
core radii after a time comparable to the half mass relaxation time
$\sim 10^9$ yrs.  Thus, their signature may last longer than 
the hardening process of the binary IMBH, but shorter than
the cluster lifetime \footnote{The binary IMBH keeps hardening at a
lower peace when the separation falls below one astronomical unit:
when in this regime, stars scattering off the binary IMBH leave the
cluster being ejected with velocities far above the escape speed.}.

Convolving the statistical results of our simulations with projection
effects, we find that few hundreds stars should display signs of
supra-thermal motion via Doppler line shift or proper
motion. Considering that only a fraction of these stars will be
detectable, the remaining being white dwarfs, neutron stars or faint
stars, a statistically significant identification of such
non-equilibrium stars seems to be very difficult.  Angular momentum
alignment requires a rather massive binary composed of two large BHs, 
for transfer to be effective. Thus, angular momentum alignment,
induced by a binary IMBH, may remain visible over a time comparable to
the cluster lifetime only if a new binary IMBH form via dynamical
processes involving other BHs after every coalescence of the
progenitor binary. This would generate families of stars with
different orientation angles, and characteristic lifetimes, each for
every binary that has formed.

Given the perturbative action that an IMBH has on stars, studies on the
overall globular cluster evolution as those from Baumgardt et
al. \cite{BAU04} can shed further light into the equilibrium properties
that a cluster with an IMBH displays.

\begin{figure}
  \includegraphics[height=.3\textheight]{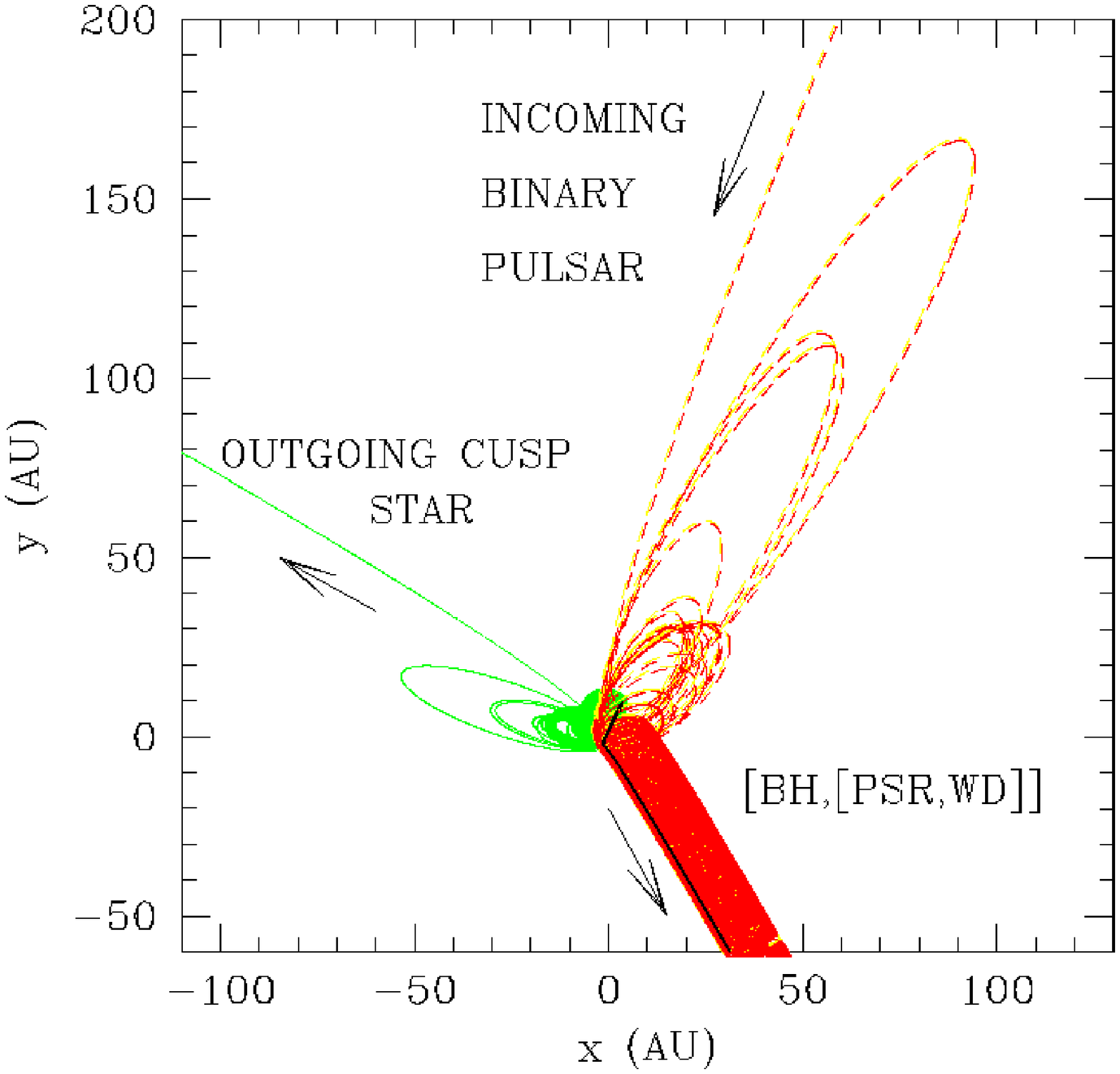}	
\includegraphics[height=.3\textheight]{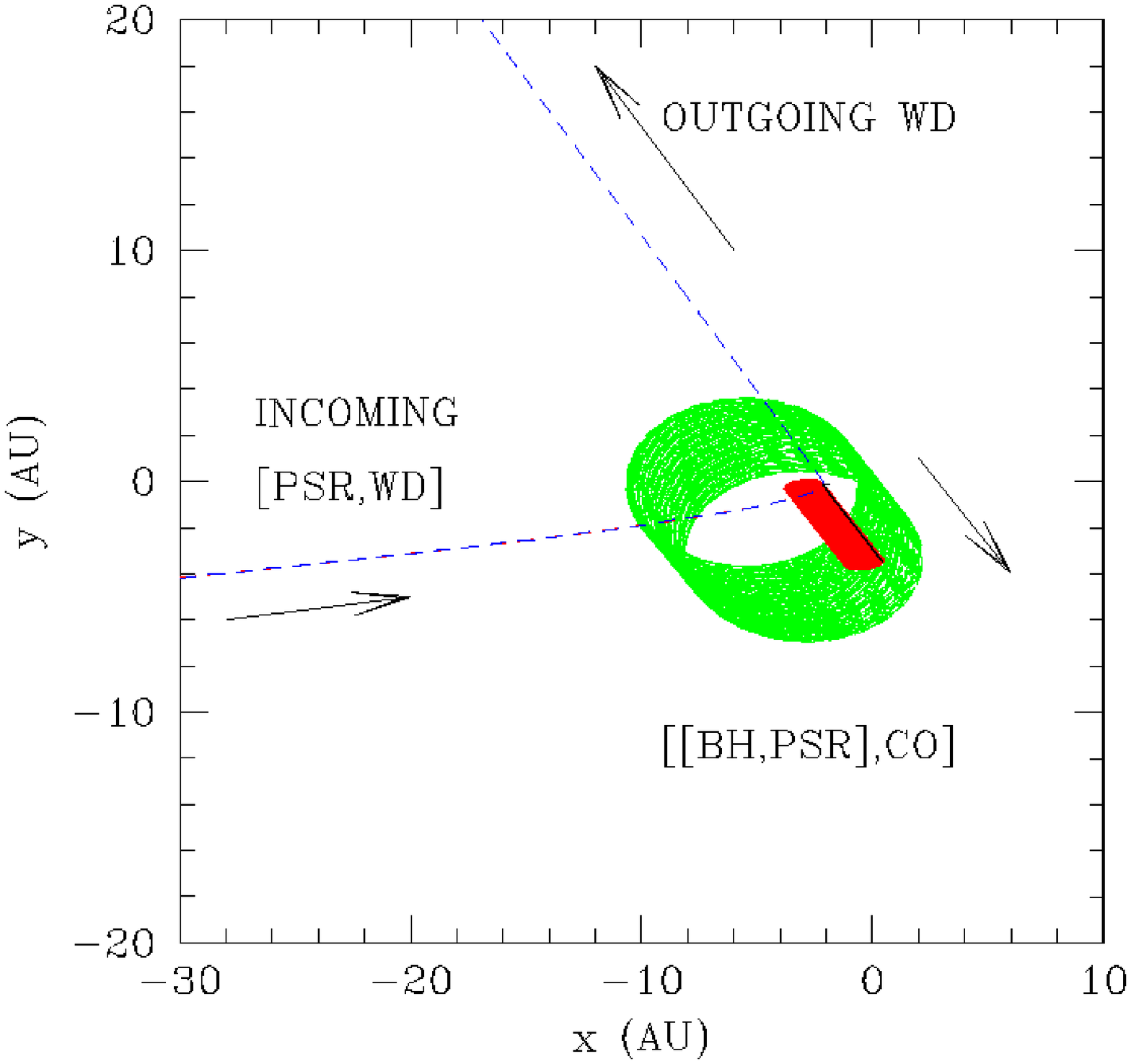}
  \caption{A stable \cite{Mar01} triplet (left panel)  
resulting from the exchange
of the cusp
star of $1\msun$ revolving around a 100$\msun$ IMBH with a binary MSP
(as PSR-A in NGC 6752), and a triplet (right panel) 
resulting from the ionization of the binary MSP impinging 
onto a 1000$\msun$ IMBH
and its cusp star. The coalescence timescale due to 
emission of gravitational waves for the inner
 binary in the first 
triplet is $1.7 \times 10^{11} $yr, while the BH-PSR binary in the second 
system should coalesce in $ 6.4\times 10^7 $yr.}
\end{figure}

\subsection{Millisecond pulsars around IMBHs}

As discuss in CMP03, binary IMBH are catalysts for the formation of
triplets, resulting from binary-binary interactions.
The encounter of a binary pulsar with a binary IMBH can create an 
extraordinary system: a millisecond pulsar-IMBH-BH  triplet (see CMP03) 
or a star-pulsar-IMBH hierarchical system.
Here we focus on an IMBH with a cusp star. 
Figure 4 shows the formation of a stable triplet through an exchange (left) and
ionization of a binary MSP (right).

We are now performing an extensive analysis of binary-single and
binary-binary encounters with single and binary IMBH to determine the
rate of formation, destruction and coalescence of these systems.  The
dynamical capture often deliver the captured pulsar (or compact object
in general) on a rather tight and eccentric orbit: by looking at the
distributions of separation and eccentricities of the triplets found
we will be able to determine also the rate of neutron star 
coalescence by gravitational waves onto the IMBH, in our nearby
universe.  These events could have a profound impact for the
gravitational waves astronomy.

\medskip
\noindent
{\bf{Acknowledgments}}
We are grateful to Stein Sigurdsson for enlightening discussion. 
This work was supported in part by the grant PRIN03-MIUR.

\bibliographystyle{aipproc}   

\IfFileExists{\jobname.bbl}{}
 {\typeout{}
  \typeout{******************************************}
  \typeout{** Please run "bibtex \jobname" to optain}
  \typeout{** the bibliography and then re-run LaTeX}
  \typeout{** twice to fix the references!}
  \typeout{******************************************}
  \typeout{}
 }

\bibliography{sample}

\IfFileExists{\jobname.bbl}{}
 {\typeout{}
  \typeout{******************************************}
  \typeout{** Please run "bibtex \jobname" to optain}
  \typeout{** the bibliography and then re-run LaTeX}
  \typeout{** twice to fix the references!}
  \typeout{******************************************}
  \typeout{}
 }

\end{document}